\documentclass{ws-procs9x6}%
\usepackage{graphicx}
\usepackage{amsmath}
\usepackage{amsfonts}
\usepackage{amssymb}%
\begin{document}
%
\title
{FAST FOURIER TRANSFORM ENSEMBLE KALMAN FILTER WITH APPLICATION TO A~COUPLED
ATMOSPHERE-WILDLAND FIRE MODEL}
\author{JAN MANDEL$^{*}$, JONATHAN D. BEEZLEY, VOLODYMYR Y. KONDRATENKO}
\address{Department of Mathematical and Statistical Sciences\\
University of Colorado Denver,
Denver, CO 80217-3364, USA\\
$^*$E-mail: Jan.Mandel@ucdenver.edu
}
\bodymatter\begin{abstract}
We propose a new type of the Ensemble Kalman Filter (EnKF), which uses the
Fast Fourier Transform (FFT) for covariance estimation from a very small ensemble
with automatic tapering,
and for a~fast computation of the analysis ensemble by convolution, avoiding the need to
solve a~sparse system with the tapered matrix. The method is combined with the
morphing EnKF to enable the correction of position errors, in addition to amplitude errors,
and demonstrated on WRF-Fire, the Weather Research Forecasting (WRF)
model coupled with  a~fire spread model implemented by the level set method.
\end{abstract}%

\section{Introduction}

Data assimilation is a statistical technique to modify the state of a running
model in response to data, based on sequential Bayesian estimation
\cite{Kalnay-AMD-2003}. The EnKF \cite{Evensen-2009-DAE} accesses the model
only as a black box. It is suitable for a wide range of problems and
relatively easy to combine with existing modeling software. However, the
ensemble size needed can be large, and the amount of computation required in
EnKF can be significant when the EnKF is modified to suppress spurious
long-range correlations. We propose a new variant of EnKF that can overcome
these difficulties by the use of FFT. The new method is applied to the
morphing EnKF \cite{Beezley-2008-MEK}, which is needed for the fire
application, but the FFT\ EnKF can be used with or without morphing.%



\section{Data Assimilation}

\label{sec:fftenkf}

We first briefly describe the EnKF for reference. The EnKF advances in time an ensemble
of simulations $u_{1},\ldots,u_{N}$, which approximates the probability
distribution of the model state $u$. The simulations are then advanced in time
until the \emph{analysis time}, when new data $d$ arrives. It is assumed that
the data error is known, $d-Hu\sim N\left(  0,R\right)  $ given $u$, where $H$
is the \emph{observation operator} and $R$ is the \emph{data error covariance
matrix}. The ensemble, now called the \emph{forecast ensemble}, is combined
with the data to give the \emph{analysis ensemble} by the EnKF formulas
\cite{Burgers-1998-ASE}%
\begin{equation}
u_{k}^{a}=u_{k}+C_{N}H^{T}\left(  HC_{N}H^{T}+R\right)  ^{-1}\left(
d+e_{k}-Hu_{k}^{f}\right)  , \label{eq:enkf}%
\end{equation}
where $C_{N}$ is an estimate of the covariance of the model state, and $e_{k}$
is a random perturbation $e_{k}\sim N\left(  0,R\right)  $. The analysis
ensemble $u_{1}^{a},\ldots,u_{N}^{a}$ is then used as the initial condition
for a new simulation, advanced to a new analysis time, and the \emph{analysis
cycle} repeats.

In the standard EnKF \cite{Burgers-1998-ASE}, $C_{N}$ is the sample covariance
computed from the ensemble. Under suitable assumptions, it can be proved that
the ensemble converges for large $N$ to a sample from the Kalman filtering
distribution and $C_{N}$ converges to the true covariance
\cite{Mandel-2009-CEK}.

\subsection{FFT EnKF and Covariance Estimation by FFT}

The EnKF formulation (\ref{eq:enkf}) relies on linear algebra only and it is
oblivious to any structure of the model state. Large ensembles (30-100 and
more) are often needed \cite{Evensen-2009-DAE}. We are interested in obtaining
a reasonable approximation of the covariance matrix from a very small sample.
For this, we take advantage of the fact that the simulation state $u$ is a
block vector, where the blocks are values of the modelled physical fields on
grids of points in a spatial domain $\Omega$, which are (discrete versions of)
smooth random functions, i.e., realizations of random fields. The covariance
of a random field drops off quickly with distance, and it is often the case in
geostatistics that random fields are stationary, that is, the covariance
between two points depends on their distance vector only
\cite{Cressie-1993-SSD}. But for a small sample, the ensemble covariance is a
matrix of low rank with large off-diagonal elements even at a great distance.
Therefore, localization techniques are used, such as tapering, which consists
of multiplication of the terms of the sample covariance by a fixed function to
force the dropoff of the covariance away from the diagonal, resulting in a
more accurate approximation of covariance for small samples
\cite{Furrer-2007-EHP}. However, solving a system with the resulting
approximate covariance matrix is expensive, because efficient dense linear
algebra, which relies on the representation of the sample covariance matrix as
the product of two rectangular matrices \cite{Mandel-2009-DAW}, can no longer
be used.

For simplicity, we explain the FFT EnKF in the 1D case. Higher-dimensional
cases work exactly the same. Also, we consider first the case when the model
state consists of one block only.

The basic operation in the EnKF (\ref{eq:enkf}) is the multiplication
$v=C_{N}u$ of a vector $u$ by an approximate covariance matrix $C_{N}$. Denote
by $u\left(  x_{i}\right)  $ the entry of vector $u$, corresponding to node
$x_{i}$ and suppose for the moment that the random field is stationary, that
is, its covariance matrix satisfies $C\left(  x_{i},x_{j}\right)  =c\left(
x_{i}-x_{j}\right)  $ for some covariance function $c$. Then $v$ is the
convolution%
\[
v\left(  x_{i}\right)  =\sum_{j}C\left(  x_{i},x_{j}\right)  u\left(
x_{j}\right)  =\sum_{j}u\left(  x_{j}\right)  c\left(  x_{i}-x_{j}\right)  .
\]
The discrete Fourier transform changes convolution to multiplication entry by
entry. Hence, if the random field is stationary, multiplication by its
covariance matrix becomes multiplication by a diagonal matrix in the frequency domain.

The proposed method of covariance estimation consists of computing the sample
covariance of the ensemble in the frequency domain, and neglecting all of its
off-diagonal terms. This is justified by the assumption that the covariance
depends mainly on the distance.

\begin{enumerate}
\item Given ensemble $\left[  u_{k}\right]  $, apply a FFT operator $F$ to
each member to obtain the member $\widehat{u}_{k}=Fu_{k}$ in the frequency domain.

\item Compute the approximate forecast covariance matrix $\widehat{C}_{N}$ in
the frequency domain as the diagonal matrix with the diagonal entries
$\widehat{c}_{i}$ equal to the diagonal entries of the sample covariance of
the ensemble $\left[  \widehat{u}_{k}\right]  $,%
\begin{equation}
\widehat{c}_{i}=\frac{1}{N-1}{\sum_{k=1}^{N}}\left\vert \widehat{u}%
_{ik}-\overline{\widehat{u}}_{i}\right\vert ^{2},\quad\overline{\widehat{u}%
}_{i}=\frac{1}{N}\sum_{k=1}^{N}\widehat{u}_{ik}. \label{eq:dii}%
\end{equation}

\end{enumerate}

Multiplication by the approximate covariance matrix $C_{N}$ then becomes in
the frequency domain
\[
u=C_{N}v\iff\widehat{u}=Fu,\quad\widehat{v}=\widehat{c}\bullet\widehat
{u},\quad v=F^{-1}\widehat{v},
\]
where $\bullet$ is entry-by-entry multiplication, $\left(  \widehat{c}%
\bullet\widehat{u}\right)  _{i}=\widehat{c}_{i}\widehat{u}_{i}$. In the
important case $H=I$ and $R=rI$ (the whole state is observed and the data
errors are uncorrelated and the same at every point), considered here, the
EnKF (\ref{eq:enkf}) in the frequency domain reduces to entry-by-entry
operations,%
\begin{equation}
\widehat{u}_{k}^{a}=\widehat{u}_{k}+\widehat{c}\bullet\left(  \widehat
{c}+r\right)  ^{-1}\bullet\left(  \widehat{d}+\widehat{e}_{k}-\widehat{u}%
_{k}^{f}\right)  . \label{eq:spectral-enkf}%
\end{equation}

In an application, the state has multiple variables. The state vector, its
covariance, and the observation matrix then have the block form%
\begin{equation}
u=\left[
\begin{matrix}
u^{\left(  1\right)  }\\
\vdots\\
u^{\left(  n\right)  }%
\end{matrix}
\right]  ,\quad C=\left[
\begin{matrix}
C^{\left(  11\right)  } & \cdots & C^{\left(  1M\right)  }\\
\vdots & \ddots & \vdots\\
C^{\left(  M1\right)  } & \cdots & C^{\left(  MM\right)  }%
\end{matrix}
\right]  ,\quad H=\left[
\begin{matrix}
H^{\left(  1\right)  } & \cdots & H^{\left(  M\right)  }%
\end{matrix}
\right]  . \label{eq:block-enkf}%
\end{equation}
Assume that the first variable is observed, then $H^{\left(  1\right)  }=I$,
$H^{\left(  2\right)  }=0$,\ldots, $H^{\left(  M\right)  }=0$. The EnKF
(\ref{eq:enkf}) then simplifies to%
\begin{equation}
u_{k}^{\left(  j\right)  ,a}=u_{k}^{\left(  j\right)  }+C_{N}^{\left(
j1\right)  }\left(  C_{N}^{\left(  11\right)  }+R\right)  ^{-1}\left(
d+e_{k}-u_{k}^{\left(  1\right)  }\right)  ,\quad j=1,\ldots,M,
\label{eq:multi-enkf}%
\end{equation}
which becomes in the frequency domain%
\begin{equation}
\widehat{u}_{k}^{\left(  j\right)  ,a}=\widehat{u}_{k}^{\left(  j\right)
}+\widehat{c}^{\left(  j1\right)  }\bullet\left(  \widehat{c}^{\left(
11\right)  }+r\right)  ^{-1}\bullet\left(  \widehat{d}+\widehat{e}%
_{k}-\widehat{u}_{k}\right)  , \label{eq:spectral-multi-enkf}%
\end{equation}
where the spectral cross-covariance between field $j$ and field $1$ is
approximated from
\begin{equation}
\widehat{c}_{i}^{\left(  j1\right)  }=\frac{1}{N-1}{\sum_{k=1}^{N}}\left(
\widehat{u}_{ik}^{\left(  j\right)  }-\overline{\widehat{u}}_{i}^{\left(
j\right)  }\right)  ^{\ast}\left(  \widehat{u}_{ik}^{\left(  1\right)
}-\overline{\widehat{u}}_{i}^{\left(  1\right)  }\right)  ,\quad
\overline{\widehat{u}}_{i}^{\left(  j\right)  }=\frac{1}{N}\sum_{k=1}%
^{N}\widehat{u}_{ik}^{\left(  j\right)  }. \label{eq:cov-multi-spectral}%
\end{equation}

\subsection{Morphing EnKF}

To treat position errors in addition to amplitude errors, FFT EnKF is combined
with the morphing EnKF \cite{Beezley-2008-MEK,Mandel-2009-DAW}. The method
uses an additional ensemble member $u_{N+1}$, called the \emph{reference
member}. Given an initial state $u$, the initial ensemble is given by
$u_{N+1}=u$ and%
\begin{equation}
u_{k}^{\left(  i\right)  }=\left(  u_{N+1}^{\left(  i\right)  }+r_{k}^{\left(
i\right)  }\right)  \circ\left(  I+T_{k}\right)  ,k=1,\ldots
,N,\label{eq:initial}%
\end{equation}
where $r_{k}^{\left(  i\right)  }$ are random smooth functions on $\Omega$,
$T_{k}$ are random smooth mappings $T_{k}:\Omega\rightarrow\Omega$, and
$\circ$ denotes composition. Thus, the initial ensemble has both amplitude and
position variability, and the position change is the same for all blocks.
Random smooth functions and mappings are generated by FFT as a Fourier series
with random coefficients that decay quickly with frequency.

The data $d$ is an observation of $u^{\left(  1\right)  }$, and it is expected
that it differs from the model in amplitude as well as in the position of
significant features, such as firelines. The first blocks of $u_{1}%
,\ldots,u_{N}$ and $d$ are then registered against the first block of the
reference member $u_{N+1}$. We find \emph{registration mappings} $T_{k}%
:\Omega\rightarrow\Omega$, $k=0,\ldots,N$ such that%
\[
u_{k}^{\left(  1\right)  }\approx u_{N+1}^{\left(  1\right)  }\circ\left(
I+T_{k}\right)  ,\quad T_{k}\approx0,\quad\nabla T_{k}\approx0,\quad
k=0,\ldots,N,
\]
where $d=u_{0}^{\left(  1\right)  }$.
Define the registration residuals $r_{k}^{\left(  j\right)  }=u_{k}^{\left(
j\right)  }\circ\left(  I+T_{k}\right)  ^{-1}-u_{N+1}^{\left(  j\right)  }$,
$k=0,\ldots,N$. The \emph{morphing transform }maps each ensemble member
$u_{k}$into the extended state vector
\begin{equation}
u_{k}\mapsto\widetilde{u}_{k}=M_{u_{N+1}}\left(  u_{k}\right)  =\left(
T_{k},r_{k}^{\left(  1\right)  },\ldots,r_{k}^{\left(  M\right)  }\right)  .
\label{eq:morphing-transform}%
\end{equation}
Similarly, the data becomes the extended data vector $d\mapsto\widetilde
{d}=\left(  T_{0},r_{0}^{\left(  1\right)  }\right)  $. The FFT EnKF method
(\ref{eq:spectral-multi-enkf}) is applied to the transformed ensemble
$\widetilde{u}_{1},\ldots,\widetilde{u}_{N}$ with the observation operator
given by$\left(  T,r^{\left(  1\right)  },\ldots,r^{\left(  M\right)
}\right)  \mapsto\left(  T,r^{\left(  1\right)  }\right)  $. The
cross-covariances between $x$ and $y$ components of $T$ and $r^{\left(
1\right)  }$ are neglected, so the covariance $C^{\left(  11\right)  }$ in
(\ref{eq:multi-enkf}) consists of three diagonal matrices, and
(\ref{eq:spectral-multi-enkf}) applies. The new transformed reference member
is obtained as $\widetilde{u}_{N+1}^{a}=\frac{1}{N}%
{\textstyle\sum\nolimits_{k=1}^{N}}
\widetilde{u}_{k}^{a}$ and the analysis ensemble $\widetilde{u}_{1}%
,\ldots,\widetilde{u}_{N+1}$ by the \emph{inverse morphing transform}%
\begin{equation}
u_{k}^{a,\left(  i\right)  }=M_{u_{N+1}}^{-1}\left(  \widetilde{u}_{k}%
^{a}\right)  =\left(  u_{N+1}^{\left(  i\right)  }+r_{k}^{a,\left(  i\right)
}\right)  \circ\left(  I+T_{k}^{a}\right)  ,\quad k=1,\ldots,N+1.
\label{eq:inverse-morphing}%
\end{equation}


%


\section{The Wildland Fire Model}

We only summarize the model very briefly. See Ref.~\refcite{Mandel-2009-DAW}
for further details, references, and acknowledgements.

The fire model runs on a 2D rectangular grid on the Earth surface, called the
fire grid. The model postulates fire line propagation speed as a function of
wind and terrain slope, and an exponential decay of fuel by combustion after
the ignition. The fire area is represented by a level set function as the set
of points where the level set function is negative. The level set function
satisfies a partial differential equation, which is solved numerically on the
fire grid by explicit time stepping. Ignition is achieved by setting the value
of the level set function so that it is negative in the ignition region. The
state of the fire model consists of the level set function and the ignition
time at the fire grid nodes. Other derived quantities included in the state
vector are the fuel fraction burned and the heat flux from the combustion.

The fire model is coupled with the Weather Research Forecasting (WRF)
atmospheric model \cite{WRF}, which runs on a much coarser 3D grid. In every
time step, the fire model inputs the horizontal wind velocity, and the heat and
vapor fluxes output from the fire model are inserted into the atmospheric
model. The fire model is distributed as WRF-Fire with WRF, and the latest
development version is also available from the authors.

%


\section{Computational Results}

\begin{figure}[t]
\begin{center}%
\footnotesize
\hspace*{-0.4in}
\begin{tabular}
[c]{cc}%
\includegraphics[width=2.7in]{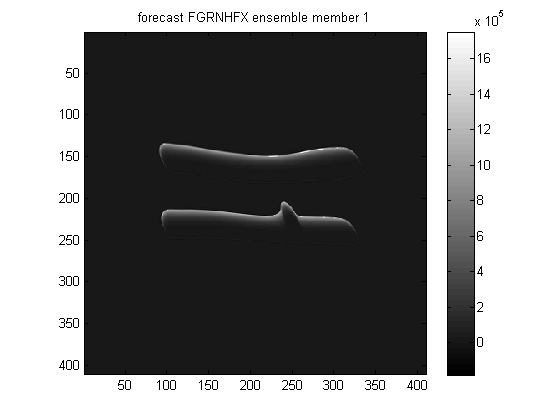} &
\hspace*{-0.4in}\includegraphics[width=2.7in]{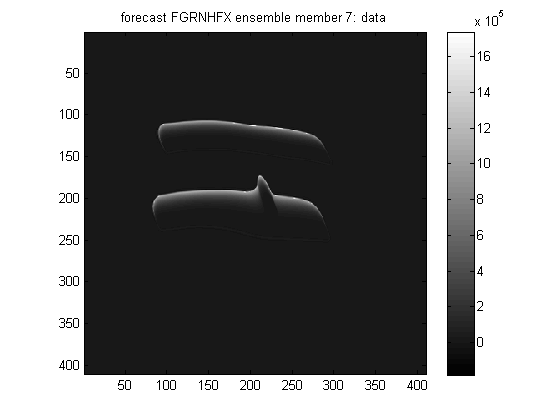}\\
{(a) Forecast member 1} & \hspace*{-0.4in}{(b) Data}\\ \ \\
\includegraphics[width=2.7in]{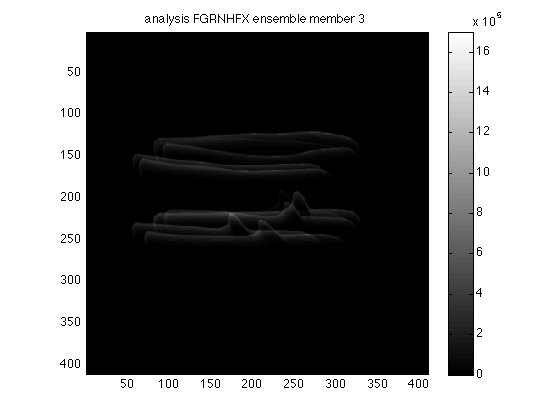} &
\hspace*{-0.4in}\includegraphics[width=2.7in]{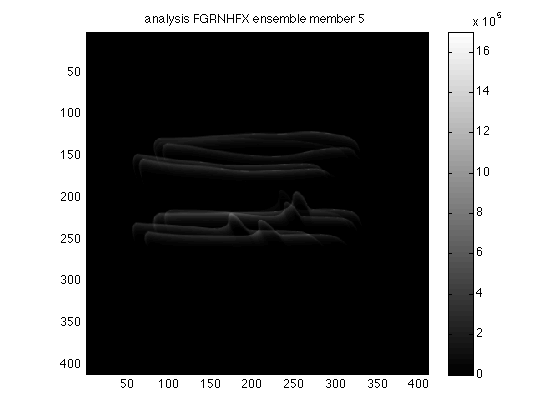} \\
{(c) Analysis  member 3} & \hspace*{-0.4in}{(d) Analysis member 5} %
\end{tabular}
\end{center}
\caption{The morphing EnKF with 5 ensemble 
members, applied  to the ground heat flux from the WRF-Fire model.
The ensemble size is not sufficient, the correct analysis is not even
approximately in the span of the forecast, and the EnKF cannot reach it.
}
\label{fig:enkf}%
\end{figure}

\begin{figure}[t]
\begin{center}%
\footnotesize
\hspace*{-0.4in}
\begin{tabular}
[c]{cc}%

\includegraphics[width=2.7in]{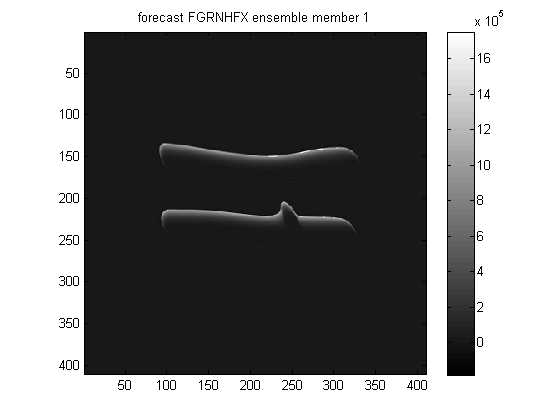} &
\hspace*{-0.4in}\includegraphics[width=2.7in]{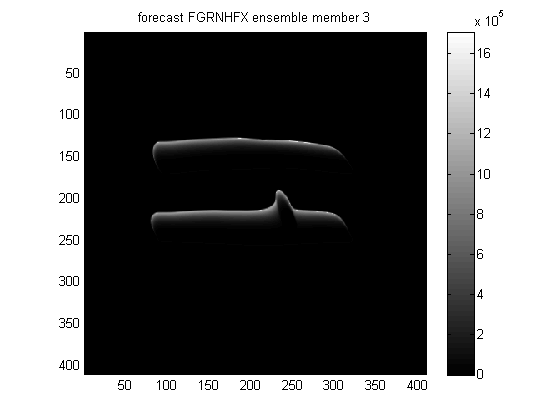}\\ 
{(a) Forecast member 1} & \hspace*{-0.4in}{(b) Forecast member 3}\\ \ \\
\includegraphics[width=2.7in]{figs/data_mem_5.png} &
\hspace*{-0.4in}\includegraphics[width=2.7in]{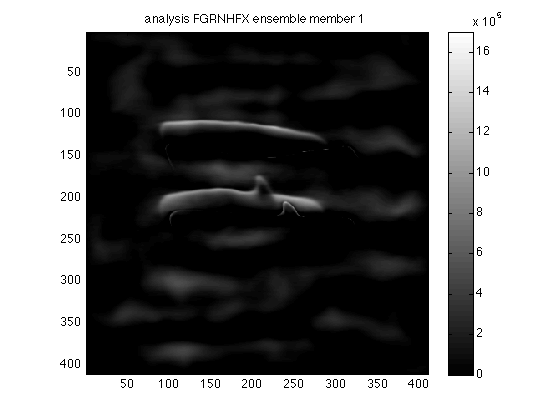}\\
{(c) Data} & \hspace*{-0.4in}{(d) Analysis member 1}\\ \ \\
\includegraphics[width=2.7in]{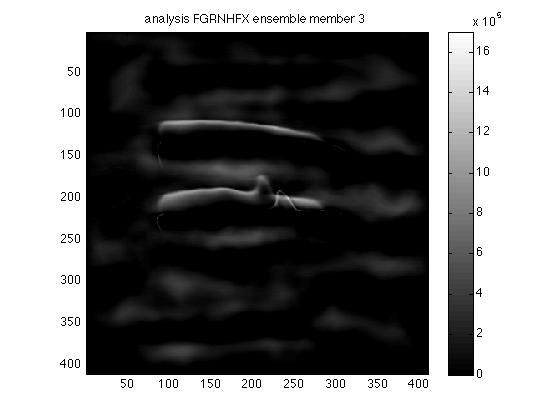}&
\hspace*{-0.4in}\includegraphics[width=2.7in]{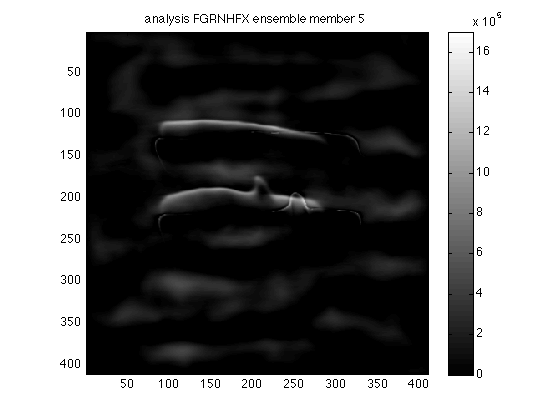}\\
{(e) Analysis  member 3} & \hspace*{-0.4in}{(f) Analysis member 5}%
\end{tabular}
\end{center}
\caption{The morphing FFT EnKF with 5 ensemble 
members, applied  to the ground heat flux from the WRF-Fire model.
The analysis ensemble moved towards the data.}
\label{fig:fftenkf}%
\end{figure}

We have used the optimization method from Ref.~\refcite{Beezley-2008-MEK} for
registration.  We have used the real $\sin$ FFT, which 
forces zero change on the boundary.
We have used a standard ideal problem distributed with WRF-Fire.
The model has a $420\times420$ fire mesh and a $42\times 42\times 41$ atmospheric
mesh. The fuel was the same on the whole domain.
The model was initialized with the wind blowing diagonally across 
the grid,
and two line ignitions and one circle ignition occur within the first 4 seconds
of simulation time.
After one minute of simulation time, when the fire was established and one of the
line fires has merged with the circular fire, the simulation was stopped and an
initial ensemble was generated by random smooth perturbation both in position
and in amplitude. 
Artificial data was created by a similar perturbation. 
The forecast was taken the same as the initial ensemble.
The 
described data assimilation algorithm was then applied with 5 members,
with the results shown in Fig.~\ref{fig:enkf} for the morphing EnKF
and Fig.~\ref{fig:fftenkf} for the morphing FFT EnKF. We see that the EnKF was not able
to approach that data at all with such a small ensemble, while the FFT EnKF 
delivered an ensemble around the correct data shape.

\section{Conclusion}

We have shown that the morphing FFT EnKF is capable of data assimilation in a 
wildfire simulation, which exhibits sharp boundaries and coherent features.
We have shown that the FFT EnKF can deliver
acceptable results with a~very small ensemble (5 members), 
unlike the standard EnKF, which is known to work with morphing 
for this application, but only with a~much larger ensemble \cite{Beezley-2008-MEK}.
Further development, involving multiple variables and multiple analysis cycles,
will be presented elsewhere.


\section{Acknowledgements}

This work was supported by NSF grants CNS-0719641 and ATM-0835579.

\bibliographystyle{ws-procs9x6}
\bibliography{ms10}

\end{document}